\documentstyle[12pt,epsfig]{article}

\def\beq{\begin{equation}}
\def\eeq#1{\label{#1}\end{equation}}
\def\eeqn{\end{equation}}
\def\beqa{\begin{eqnarray}}
\def\eeqa#1{\label{#1}\end{eqnarray}}
\def\eeqan{\end{eqnarray}}
\def\CR{\nonumber \\ }
\def\leqn#1{(\ref{#1})}

\def\to{\rightarrow}

\def\stacksymbols #1#2#3#4{\def\theguybelow{#2}
    \def\vp{\lower#3pt}
    \def\sp{\baselineskip0pt\lineskip#4pt}
    \mathrel{\mathpalette\intermediary#1}}

\def\intermediary#1#2{\vp\vbox{\sp
     \everycr={}\tabskip0pt
     \halign{$\mathsurround0pt#1\hfil##\hfil$\crcr#2\crcr
              \theguybelow\crcr}}}

\begin{document}

\begin{titlepage}

\vskip 5cm
\begin{center}
{\Huge \bf Antiprotons from Dark Matter: }
\vskip0.5cm
{\Huge \bf Effects of a Position-Dependent } 
\vskip0.4cm
{\Huge \bf Diffusion Coefficient} 

\vskip.2cm
\end{center}
\vskip1cm

\begin{center}
{\bf Maxim Perelstein and Bibhushan Shakya} \\
\end{center}
\vskip 8pt

\begin{center}
	{\it Institute for High Energy Phenomenology\\ 
	Newman Laboratory of Elementary Particle Physics\\
	     Cornell University, Ithaca, NY 14853, USA } \\

\vspace*{0.3cm}

{\tt  mp325@cornell.edu, bs475@cornell.edu}
\end{center}

\vglue 0.3truecm

\begin{abstract}
\noindent Energetic antiprotons in cosmic rays can serve as an important indirect signature of dark matter. Conventionally, the antiproton flux from dark matter decay or annihilation is calculated by solving the transport equation with a space-independent diffusion coefficient within the diffusion zone of the galaxy. Antiproton sources outside the diffusion zone are ignored under the assumption that they propagate freely and escape. In reality, it is far more likely that the diffusion coefficient increases smoothly with distance from the disk, and the outlying part of the dark matter halo ignored in the conventional approach can be significant, containing as much as 90\% of the galactic dark matter by mass in some models. We extend the conventional approach to address these issues. We obtain analytic approximations and numerical solutions for antiproton flux for a diffusion coefficient that increases exponentially with the distance from the disk, thereby including contributions from dark matter annihilation/decay in essentially the full dark matter halo. We find that the antiproton flux predicted in this model deviates from the conventional calculation for the same dark matter parameters by up to about 25\%. 

\end{abstract}

\end{titlepage}

\section{Introduction}

While astronomical observations have firmly established that about 80\% of the matter content of the universe exists in the form of non-baryonic dark matter, its microscopic properties remain hitherto unknown. 
Weakly interacting massive particles (WIMPs) are the best-motivated candidates for dark matter
from a theoretical point of view. WIMPs couple weakly to standard model (SM) particles, opening the possibility of dark matter particles decaying or annihilating into SM final states. Energetic cosmic rays produced in such decay or annihilation processes -- in particular antimatter, which is rarely produced in astrophysical processes -- can serve as important indirect signatures of dark matter in the galaxy. Observation of such signals can reveal information on the microscopic properties of dark matter. 

Our focus in this paper is on antiprotons produced in the annihilation or decay of dark matter in the Milky Way galaxy. This has become a topic of significant interest following recent measurements of  the antiproton flux and the ratio of antiproton to proton flux up to 180 GeV by the PAMELA experiment~\cite{pamela, pamela2}. No deviations from the expected astrophysical flux were observed. This data can in principle be used to put bounds on dark matter properties; however, to do this, the effects of propagation of antiprotons between the dark matter decay/annihilation location and the detector must be properly accounted for. 

Propagation of antiprotons in the galaxy is governed by transport equations, which incorporate interactions with galactic magnetic fields and the interstellar medium (ISM). Conventional calculations solve the transport equation in a two-zone model: a thin disk of $\mathcal{O}$($100$ pc) thickness on the galactic plane where the interactions with the ISM occur, embedded in a larger diffusion region where galactic magnetic fields are appreciable and trap cosmic rays. The diffusion region is taken to be a cylinder of radius $R$ of order 20 kpc and half-thickness $L$ of order 1--10 kpc, and the diffusion coefficient is assumed to be position-independent inside this region. Outside this diffusion region, antiprotons are assumed to propagate freely, leading to vanishing antiproton density on the boundaries of the cylinder in the steady state.  With these assumptions, the transport equations can be solved in a straightforward manner. In particular, for antiproton energies of interest for dark matter searches (of order 10 GeV and above), additional approximations can be made that allow an {\it analytic} solution to the antiproton density and flux in the form of a Bessel series~\cite{solutions}. 

The conventional two-zone model is, however, a rather crude approximation. First, the assumption of a sharp boundary at $L$ between the diffusion and free-propagation zones is unphysical. In reality, magnetic fields are known to decrease gradually with the distance from the disk, with exponential decay providing a reasonable fit to data. The diffusion coefficient likely follows a similar exponential profile~\cite{exponentialk}. Second, a typical dark matter halo is spherically symmetric and extends {\it beyond} the diffusion region and into the free propagation zone, in particular in the vertical direction: for example, for an isothermal dark matter profile in a model with $L=1$ kpc, the diffusion zone contains only 10\% of the dark matter mass of the full halo. The two-zone model completely ignores the antiprotons produced by dark matter decay/annihilation outside the diffusion zone.

The aim of this paper is to extend the conventional formalism for antiproton flux calculations to overcome these shortcomings.\footnote{The extension presented here is similar to the formalism developed by us in a previous paper~\cite{positrons} for positrons. However, the absence of energy-loss terms for antiprotons allows us to make significant progress analytically, which was not possible for positrons.} After reviewing the conventional formalism in Sec.~\ref{sec:oldform}, we consider a three-zone model in 
Sec.~\ref{sec:newform1}. This model still assumes an abrupt change in the diffusion coefficient at $L$, but includes the sources in the free-propagation zone extending to $D\gg L$, so that essentially all of the antiprotons from the dark matter halo are taken into account. We find an analytic solution to this model, and show that in the limit when diffusion in the free-propagation zone is completely absent, the sources in this zone have no effect on the flux measured at Earth. We then consider a model with an exponentially varying diffusion coefficient~\cite{exponentialk} in Sec.~\ref{sec:newform2}, and obtain both a numerical solution and two analytic approximations which converge to it at high energies (above 50 GeV or so, depending on the desired accuracy). In Sec.~\ref{sec:results}, we present numerical results quantifying the effects of this more realistic propagation model on the antiproton fluxes produced by dark matter decay or annihilation, and the corresponding bounds on dark matter properties. We close with concluding remarks in Sec.~\ref{sec:conc}.

\section{Antiproton Flux: The Conventional Formalism}
\label{sec:oldform}

The starting point for conventional calculations is the steady-state diffusion equation for antiprotons \cite{solutions}:
\beq
-\, \nabla\,\left[ K({\bf x}, E)\,\nabla n_{\bar{p}} \right]\,+\frac{\partial}{\partial z}(V_C(z)n_{\bar{p}}(E,{\bf x}))+2h\delta(z)\Gamma_{ann}n_{\bar{p}}(E,{\bf x})=\, q_{\bar{p}}({\bf x}, E)\,,
\eeq{transport}
where $n_{\bar{p}} $ is the antiproton density, $K$ is the diffusion coefficient, the convective wind term
\beq
V_c(z) \,=~{\rm sign}\, (z) \,V_C 
\eeq{wind}
is representative of motion of the medium responsible for diffusion, and $q$ is the source term describing antiproton production in dark matter annihilation or decay.\footnote{The source also contains the contribution of the so-called ``tertiary" term, describing inelastic antiproton collisions with the ISM. 
We will ignore this contribution throughout this paper, as it is expected to be numerically unimportant for dark matter studies. In principle, it could be included in the calculations using an iterative approach.}    The $2h\delta(z)\Gamma_{ann}$ term corresponds to antiproton interactions with the Interstellar Medium (ISM); $h=0.1$~kpc is the half-width of the interaction zone, and $\Gamma_{ann}$ is the annihilation rate between antiprotons and protons, parameterized as~\cite{winodm}
\begin{equation}
\Gamma_{ann}=(n_H+4^{2/3}n_{He})\sigma^{ann}_{\bar{p}p}v_{\bar{p}}\,,
\end{equation}
where $n_H\sim1$ cm$^{-3}$ and $n_{He}\sim0.07n_H$ are the hydrogen and helium number densities, and $\sigma^{ann}_{\bar{p}p}(T)$ is given by~\cite{annterm2, fragmentation}
\beq
\sigma^{{\rm ann}}_{\bar{p}p}(T)= \cases{661(1+0.0115(T/{\rm GeV})^{-0.774}-0.948 (T/{\rm GeV})^{0.0151}){\rm ~mb}, & $T < 15.5$~{\rm GeV}; \cr\cr
36(T/{\rm GeV})^{-0.5}~{\rm mb}, & ~{\rm otherwise.}}
\eeqn
Energy loss due to bremsstrahlung and collisions with cosmic microwave background photons, an important aspect of positron propagation, is a negligible effect for the more massive antiprotons and hence absent in the diffusion equation~\leqn{transport}. Following~\cite{solutions, dmseeslight}, we have also ignored solar modulation effects, which are expected to be unimportant at the high $\bar{p}$ energies (10 GeV and above) of interest for dark matter searches. 

The conventional approach is to assume an energy dependent but position independent diffusion coefficient
\beq
K(E)=K_0\beta({\mathcal{R}}_{{\rm GV}})^\delta\,,
\eeq{diffcoeff}
where $\beta=v/c$ and ${\mathcal{R}}_{{\rm GV}}=pc/eZ$ is the rigidity of the particle measured in gigavolts (GV). Eq.~\leqn{transport} is then solved in a cylindrical region of radius $R$ and half-thickness $L$, with vanishing antiproton density at the boundaries of the cylinder. The parameters $L, R,K_0,V_C$ and $\delta$ define the galactic propagation model. We list three commonly used models in Table~\ref{tab:galprop}; these values are obtained from the analysis of observed isotope ratios in cosmic rays, primarily boron to carbon (B/C) ratio~\cite{BCratio}. 

\begin{table}[t]
\begin{center}
\begin{tabular}{|c|c|c|c|c|}
\hline
Model & $\delta$ & $K_0$ (kpc$^2$/Myr) & $L$ (kpc) & $V_C$ (km/s)\\
\hline
MIN & 0.85 & 0.0016 & 1 & 13.5\\
MED &0.70 &0.0112 &4 & 12\\
MAX &0.46 &0.0765 &15 & 5\\
\hline
\end{tabular}
\caption{Galactic propagation models~\cite{propmodel, BCratio}. All models set $R=20$ kpc.}
\label{tab:galprop}
\end{center}
\end{table}

An analytic solution can be obtained by expanding $n_{\bar{p}}$ as a Bessel series
\beq
n_{\bar{p}}(\rho,z,E)=\sum_{i} N_i(z,E)J_0\left(\frac{\zeta_i\rho}{R}\right)\,,
\eeq{bessel}
where $J_0$ is the zeroth order Bessel function of the first kind, and $\zeta_{i}$'s are the zeros of $J_0$. 
This reduces Eq.~\leqn{transport} to a set of ordinary differential equations on $N_i(z)$, with $E$ acting simply as a label. The source for each $N_i$ is given by the Bessel transform of the source term,
\beq
q_i(z) = \frac{2}{J_1(\zeta_i)^2R^2}\int^R_0 d\rho \rho J_0\left(\frac{\zeta_i\rho}{R}\right)q(\rho,z)\,,
\eeq{source_bes}
where $J_1$ is the first-order Bessel function of the first kind.
The solution has the form~\cite{solutions}
\beq
N_i(z)=e^{a(|z|-L)}\frac{y_i(L)}{B_i\sinh(S_iL/2)}\left[\cosh(S_iz/2)+A_i\sinh(S_iz/2)\right]-\frac{y_i(z)}{KS_i}\,,
\eeq{oldsol}
where we defined $a=V_C/(2K)$, as well as
\beqa
S_i&=&2 \left(a^2+\frac{\zeta_i^2}{R^2}\right)^{1/2}\,,~~~A_i = \frac{V_C+2h\Gamma_{ann}}{KS_i}; ~~~B_i\,=\, KS_i \left[ A_i + \coth(S_iL/2) \right]\,,\CR
y_i(z) &=& 2\int^z_0e^{a(z-z')}\sinh\left[S_i(z-z')/2\right]\,q_i(z')dz'\,. 
\eeqa{conv_defs}
In particular, at the position of the Earth (corresponding to $z=0$), where the fluxes are measured, the solution simplifies to
\beq
N_i(0)=\frac{e^{-aL} y_i(L)}{B_i\sinh(S_iL/2)}.
\eeq{zerosolution}
The antiproton flux at the top of the Earth's atmosphere can then be calculated as
\beq
\Phi_{\bar{p}}(E) \,=\, \frac{\beta_{\bar{p}}}{4\pi} \,n_{\bar{p}}(r_\odot, z_\odot=0, E),
\eeq{flux}
where $r_\odot=8.5$ kpc is the distance from the solar system to the galactic center.

\section{Three-Zone Propagation Model}
\label{sec:newform1}

\begin{figure}[t]
\centering
\begin{tabular}{cc}
\includegraphics[width=3in,keepaspectratio]{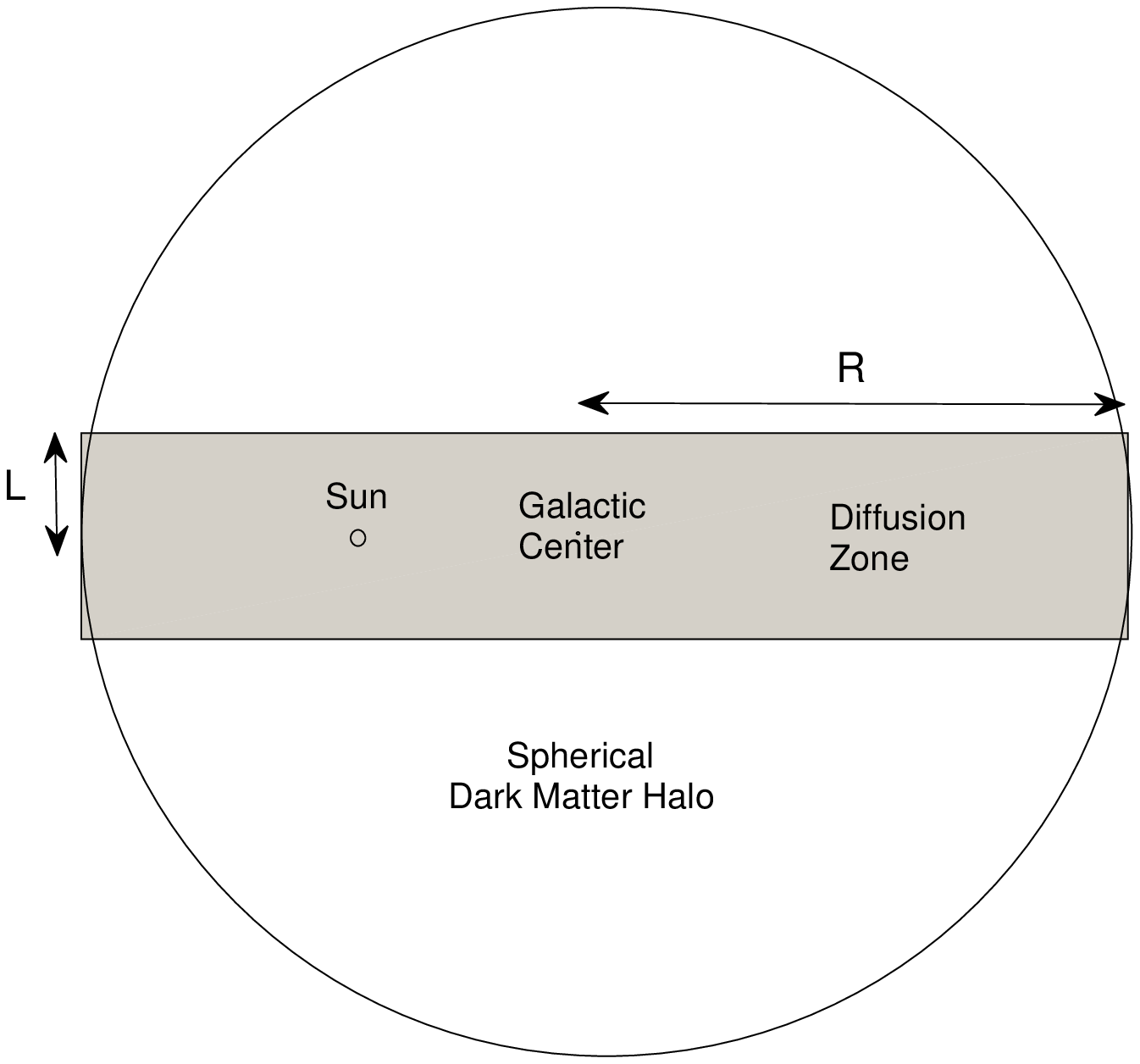}&
\includegraphics[width=3in,keepaspectratio]{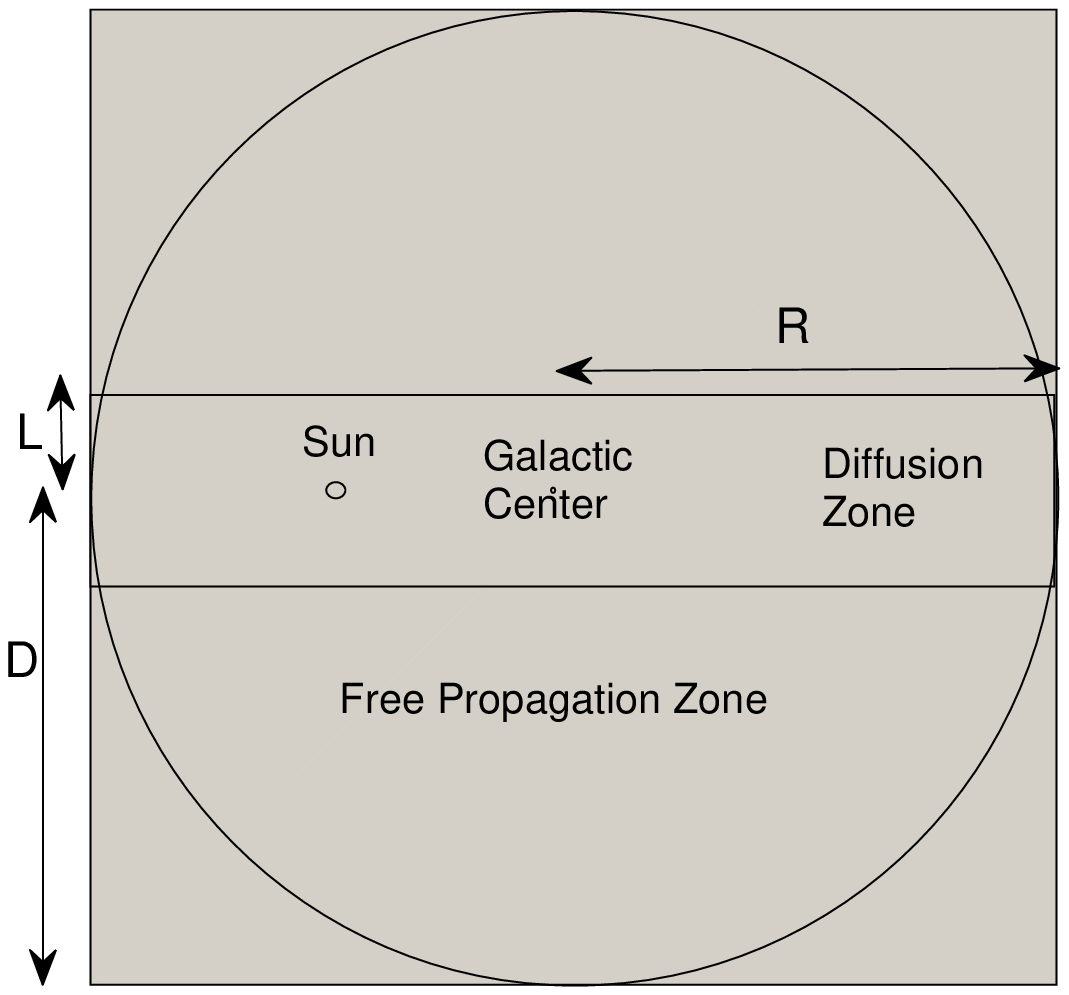}\\
\end{tabular}
\caption{Left: The dark matter halo extends significantly beyond the diffusion zone, but only sources inside the zone are considered in the conventional formalism (Section~\ref{sec:oldform}). Right: The three-zone formalism of Section~\ref{sec:newform1} includes sources in the free propagation zone in addition to the diffusion zone.}
\label{fig:geometry}
\end{figure}

A straightforward way to include contributions from sources outside of the diffusion zone is to 
consider the transport equation in a cylinder of half-thickness $D>L$, choosing $D$ sufficiently large so that all or virtually all of the dark matter halo is contained within this cylinder (see Fig.~\ref{fig:geometry}, right). The diffusion coefficient remains of the form~\leqn{diffcoeff} for $|z|\textless L$, whereas in the free propagation region $L\leq |z| \leq D$ we assume
\beq
\tilde{K}(E)=K_1\beta({\mathcal{R}}_{{\rm GV}})^\delta
\eeq{diffcoeff2}
with $K_1\gg K_0$, corresponding to significantly weaker magnetic fields and longer diffusion lengths in this region. To describe completely free propagation, one can take $K_1\to\infty$ at the end of the calculation. Note that the choice of the energy dependence in Eq.~\leqn{diffcoeff2}, which was taken to be the same as in the diffusion region, is a matter of mathematical convenience; we do not expect the results to depend on this choice in the large $K_1$ limit. We impose the boundary condition $n_{\bar{p}}=0$ on the boundaries of the extended cylinder, since the (extragalactic) sources and magnetic fields outside the cylinder can be safely neglected.

The transport equation in the three-zone approach can still be solved analytically. Expanding $n_{\bar{p}}$ in Bessel series as before, the solutions for $N_i$ inside each zone are obtained as in the conventional approach. For each $N_i$, the solution in each zone contains two free constants parametrizing the solutions of the homogeneous (sourceless) equation. Two of these constants are determined by boundary conditions ($N_i=0$ at $z=D$ and the matching condition at $z=0$, which is the same as in the conventional approach). The other two are obtained from the matching conditions at the boundary between the zones at $z=L$:
\beq
N_i(L-\epsilon) = N_i(L+\epsilon),~~~K\,\frac{dN_i}{dz}(L-\epsilon) = \tilde{K}\,\frac{dN_i}{dz}(L+\epsilon)\,.
\eeq{match}  
Physically, these conditions describe the continuity of antiproton density and flux, respectively, across the boundary; the latter condition is independent of the convection term, which is continuous at $z=L$. The resultant solution in the diffusion zone ($|z|\leq L$) has the form
\beq
N_i(z)=e^{a(|z|-L)}\frac{\alpha_i}{\beta_i}\left[\cosh(S_iz/2)+A_i\sinh(S_iz/2)\right]-\frac{y_i(z)}{KS_i}\,.
\eeq{newsol}
The numerator and denominator in the coefficient are given by
\beqa
\alpha_i &=& \tilde{y}_i(D)e^{-2\tilde{a}\Delta}\,+\,Y_i(L)\sinh(\tilde{S}_i\Delta) \,+\,
\frac{\tilde{K} \tilde{S_i}}{K S_i} \, y_i(L) \cosh(\tilde{S}_i\Delta) \,; \CR
\beta_i &=& \tilde{K} \tilde{S}_i \left[ A_i\sinh(S_iL/2)\,+\,\cosh(S_iL/2)\right] \, \cosh(\tilde{S}_i\Delta) \CR & & + KS_i\,\left[ A_i\cosh(S_iL/2)\,+\,\sinh(S_iL/2) \right] \, \sinh(\tilde{S}_i\Delta)\,.
\eeqa{alpha_beta}
Here, $S_i, A_i, a$, and $y_i$ are as defined in Section~\ref{sec:oldform}, and we also defined 
\beqa
\tilde{a}&=&\frac{V_C}{2\tilde{K}},~~~~~\tilde{S_i}=2\left(\tilde{a}^2+\frac{\zeta_i^2}{R^2}\right)^{1/2},~~~~~\Delta=(D-L)/2, \CR
Y_i(z)&=&2\int^z_0 e^{a(z-z')}\cosh (S_i(z-z')/2)\,q_i(z')\,dz', \CR
\tilde{y}_i(z) &=& 2\int^z_L e^{\tilde{a}(z-z')}\sinh(\tilde{S}_i(z-z')/2)\,q_i(z')\,dz'.
\eeqa{defs}
In particular, the flux at Earth is given by
\beq
\Phi_{\bar{p}}(E) \,=\, \frac{\beta_{\bar{p}}(E) \, e^{-aL}}{4\pi} \, \sum_i \frac{\alpha_i}{\beta_i}\, J_0 \left( \frac{\zeta_i r_\odot}{R}\right)\,.
\eeq{newsolzero}

From the above solution, it can be easily seen that in the limit $\tilde{K}\to\infty$ the solution inside the diffusion zone reduces to the one obtained in the conventional formalism, up to corrections of order $K/\tilde{K}$. In other words, the sources outside the diffusion zone {\it do not} contribute to the flux observed on Earth: all antiprotons from those sources get reflected by the boundary of the diffusion zone and do not penetrate it. This result in a sense justifies the use of conventional formalism, which ignores such sources. On the other hand, this result is only valid assuming an infinite, and infinitely sharply localized, jump in the diffusion coefficient at the boundary of the diffusion zone; both assumptions are clearly unphysical. It is far more reasonable to expect that the diffusion coefficient changes smoothly on the few-kpc length scale away from the disk. This motivates the analysis of the following section.

\section{Exponentially Increasing Diffusion Coefficient}
\label{sec:newform2}

Since diffusion is caused by charged particles getting confined by galactic magnetic fields, variations in the diffusion coefficient should follow variations in the magnetic field strength. The magnetic field in the galaxy, while not precisely known, is believed to follow the approximate profile~\cite{magneticfields}
\beq
B(\rho,z)\approx (11\mu {\rm G}) \,\times \exp \left(-\frac{\rho}{10~{\rm kpc}}-\frac{|z|}{2~{\rm kpc}}\right).
\eeq{bfield}
The diffusion coefficient is expected to have a similar exponential spatial dependence. Ignoring radial dependence, we can model the diffusion coefficient as~\cite{exponentialk}:
\beq
K(E,z)=K_e(E)\exp(|z|/z_t),~~~~~K_e(E)=K_2\beta\left(\frac{\mathcal{R}_{{\rm GV}}}{3}\right)^\delta.
\eeq{kextended}
A numerical code for propagation of cosmic rays with a diffusion coefficient of this form has been presented in Ref.~\cite{exponentialk}. Using this code, it was demonstrated that a consistent fit to observed cosmic ray isotope ratios can be obtained in this model. Typical values of the parameters producing consistent fits are as follows: $z_t=4$ kpc, $\delta=0.57$, $K_2=(0.55\times 10^{28}$~cm$^2$s$^{-1}$kpc$^{-1}) \times z_t \,=\, 2.2\times 10^{28}$ cm$^2$s$^{-1}$. We will use these values in all numerical calculations and plots in this paper. 

Our goal is to solve Eq.~\leqn{transport} with the diffusion coefficient of the form~\leqn{kextended} and a boundary condition of vanishing antiproton density at large distances away from the galaxy. In practice, we demand $n_{\bar{p}}=0$ at $|z|=L$ and $\rho=R$, choose $L$ and $R$ sufficiently large so that essentially all of the dark matter halo is contained in the cylinder $|z|\leq L$, $\rho\leq R$, and solve the equation within this cylinder. 
Since the setup still possesses cylindrical symmetry, the Bessel transform~\leqn{bessel} can be used as before to reduce the transport equation to an (infinite) set of ordinary differential equations. We were unable to find closed-form analytic solutions of these ODEs for the diffusion coefficient of the form~\leqn{kextended} and the wind term of the form~\leqn{wind}, but it is straightforward to solve them numerically, terminating the Bessel series after a finite number of terms, $N_B$. We verified that with typical dark matter profiles, the Bessel series converges rapidly, and convergence at a 1\% level is achieved for $N_B\leq 50$ in all cases we studied (see Section~\ref{sec:results}). We also find it useful to consider two situations, slightly different from the real one, in which an analytic solution {\it can} be easily found:
\begin{itemize}
\item Case 1: The convective wind term has the same exponential profile as the diffusion coefficient, namely $V_C(z) = V_e \exp(|z|/z_t)$;
\item Case 2: The convective wind term is neglected.
\end{itemize}
While both situations are unphysical, they in a sense ``bracket" the desired one (constant non-zero wind term): The first one underestimates the antiproton flux due to the artificially large wind term at large $z$, while the second one overestimates the flux. For case 1, the solution closely resembles Eq.~\leqn{oldsol}: 
\beq
N_i(z)\,=\,e^{a(|z|-L)}\,\frac{y_i(L)}{B_i\sinh(S_iL/2)}\,\left[\cosh(S_iz/2)+A_i \sinh(S_iz/2)\right]-\frac{y_i(z)}{K_eS_i}\,,
\eeq{newsolexp}
although with slightly different definitions:
\beqa
a &=& \frac{V_e}{2K_e}\,-\,\frac{1}{2z_t}\,,~~~
S_i \,=\, 2\left(a^2+\frac{\zeta_i^2}{R^2}+\frac{V_e}{z_tK_e}\right)^{1/2},\CR
A_i &=& \frac{V_e+2h\Gamma_{ann}}{K_eS_i} + \frac{1}{z_t S_i}; ~~~B_i\,=\, K_eS_i \left[ A_i + \coth(S_iL/2) \right], \CR
y_i(z) &=& 2\int^z_0\,e^{a(z-z')}\,\sinh\left[S_i(z-z')/2\right]\,q_i(z')\,e^{-z'/z_t}\,dz'.
\eeqa{more_defs}
For case 2, one can simply set $V_e=0$ in the above solution. As the wind term becomes negligible at high antiproton energies, the case 1 and 2 fluxes should approach each other, and therefore the true solution, in this regime. As we will see in the next section, the antiproton fluxes predicted in cases 1 and 2 are actually quite close to each other throughout the energy range relevant for dark matter searches.

\section{Results}
\label{sec:results}

We model the dark matter distribution in the Milky Way galaxy with two widely used profiles, isothermal and Einasto. The isothermal profile is given by
\beq
\rho(r)=\rho_\odot\frac{1+(r_\odot/r_s)^2}{1+(r/r_s)^2}\,,
\eeq{isoth}
where $\rho_\odot=0.3$ GeV cm$^{-3}$ is the local dark matter density in the solar neighborhood, and $r_s=5$~kpc. The Einasto profile is~\cite{ref:einasto}  
\beq
\rho(r)=\rho_\odot \exp \left[-\frac{2}{\alpha}\left(\frac{r^{\alpha}-r_\odot^{\alpha}}{(25~{\rm kpc})^{\alpha}}\right)\right]
\eeq{einasto}
with $\alpha=0.17$. We studied both annihilating ($q\propto \rho^2$)  and decaying ($q\propto \rho$) dark matter scenarios. 

\subsection{Flux Ratios}

\begin{figure}[t]
\centering
\begin{tabular}{cc}
\includegraphics[width=3.2in,height=2.1in]{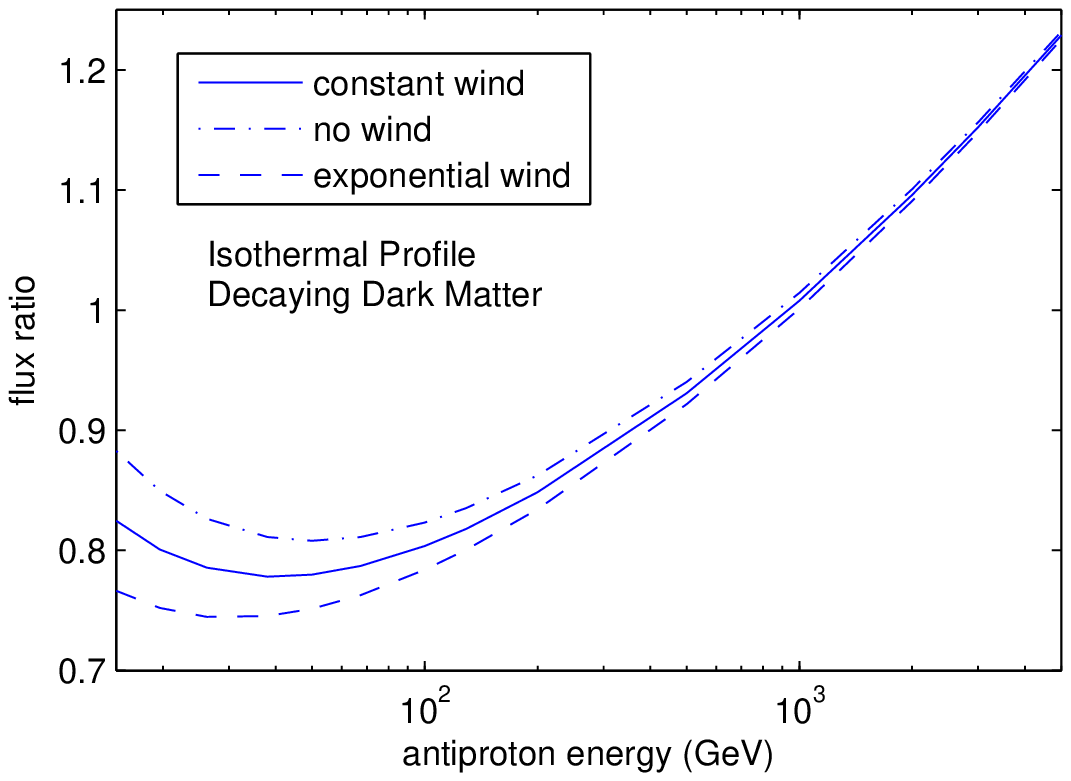}&
\includegraphics[width=3.2in,height=2.1in]{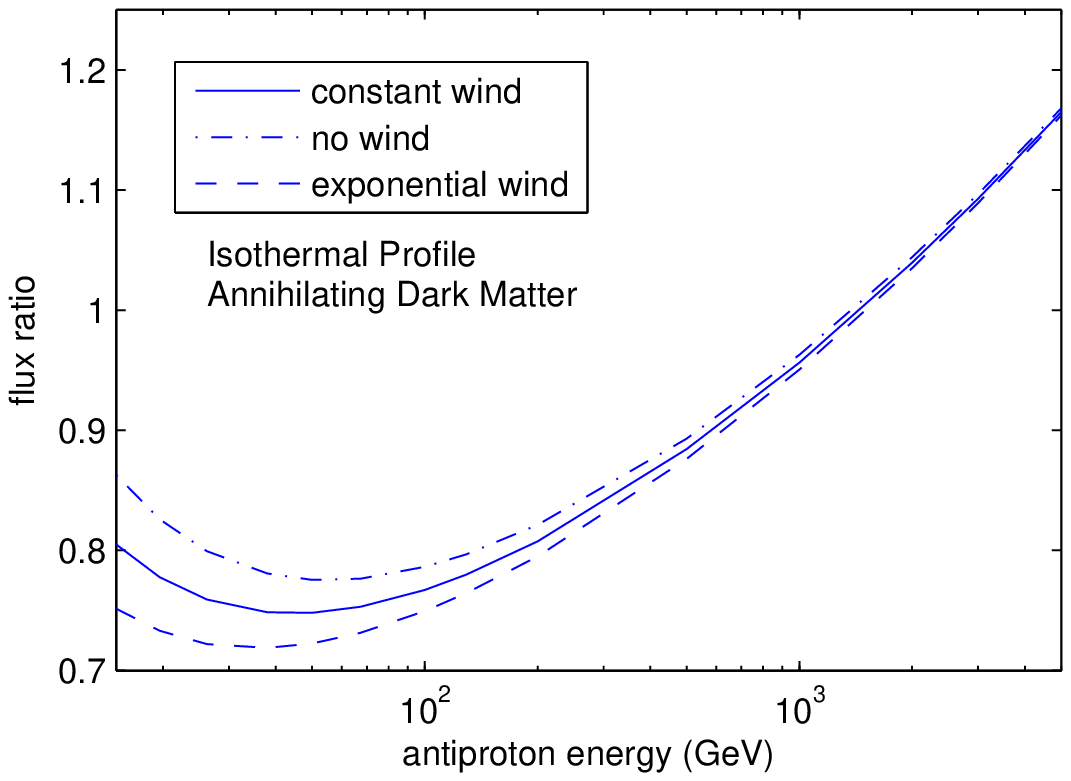}\\
\includegraphics[width=3.2in,height=2.1in]{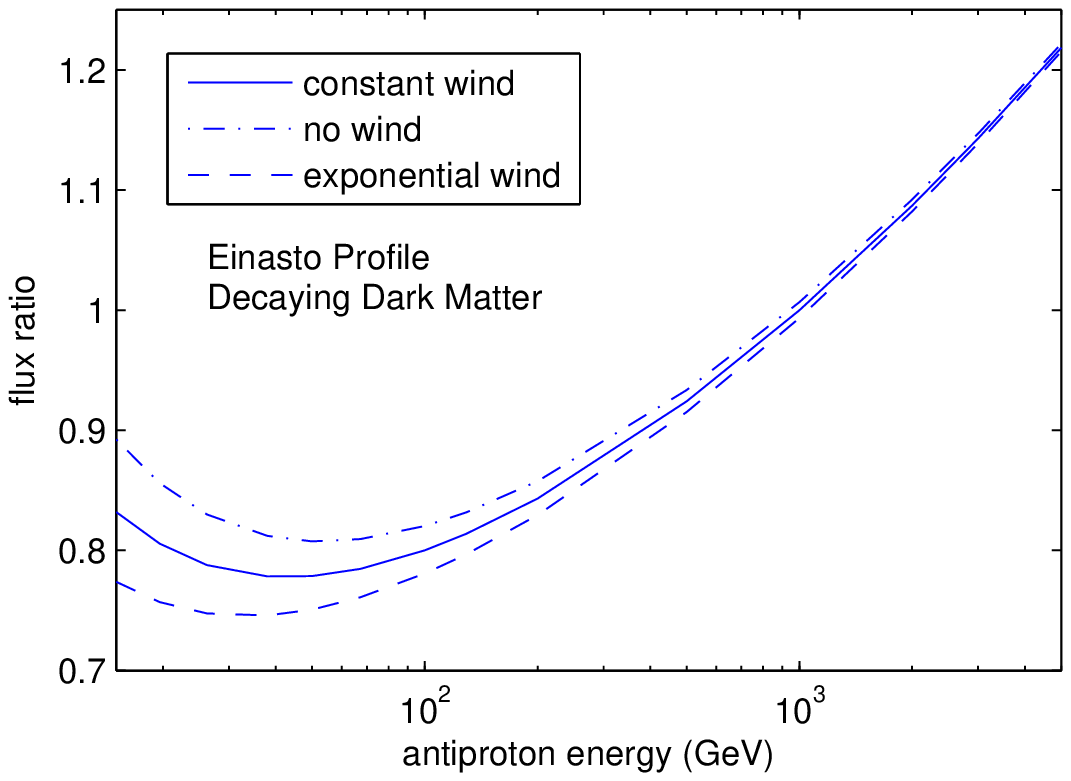}&
\includegraphics[width=3.2in,height=2.1in]{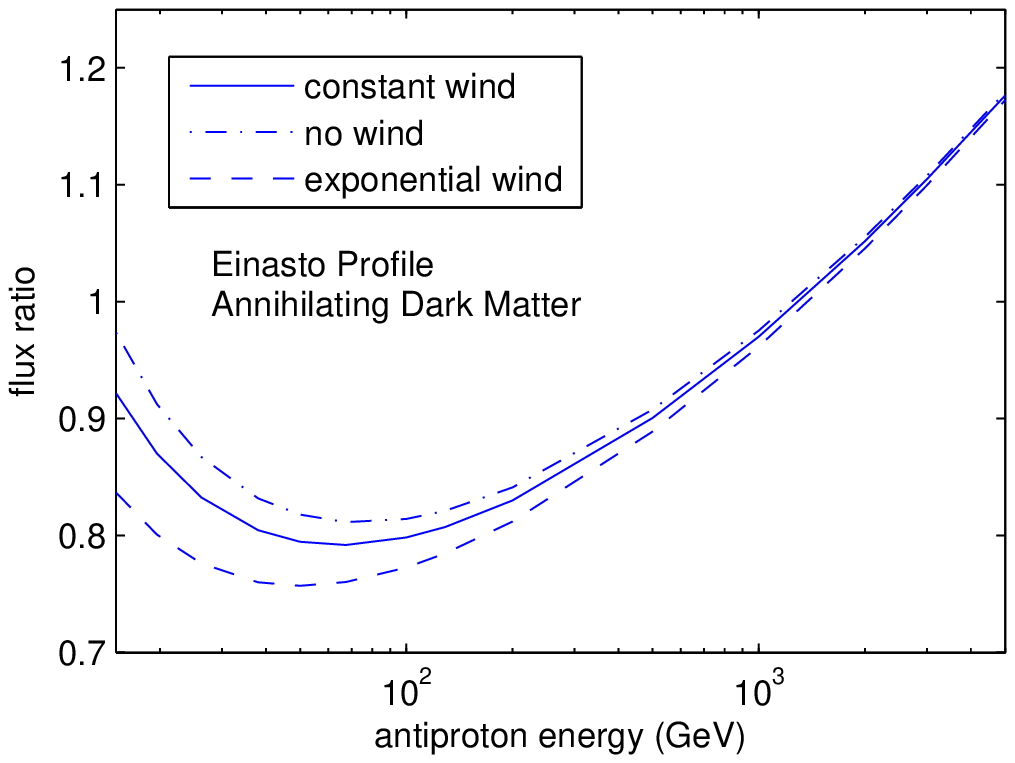}\\
\end{tabular}
\caption{Ratio of antiproton flux computed from exponential diffusion coefficient model to flux computed from conventional model for isothermal (top) and Einasto (bottom) profiles. The left (right) panels show the results for decaying (annihilating) dark matter. Dashed and dot-dashed lines correspond to fluxes analytically calculated with and without the exponential wind term respectively. The solid line in between corresponds to fluxes calculated numerically with a constant convective wind term.}
\label{fig:result1}
\end{figure}

Figure~\ref{fig:result1} shows the flux calculated with the exponential diffusion coefficient model described above, with $V_e=5$~km/s, $L=40$~kpc, $R=20$~kpc, and other parameters as specified in Sec.~\ref{sec:newform2}, as a function of antiproton energy. The flux is normalized to the flux calculated within conventional formalism (MED propagation model). Results are shown for Einasto and isothermal profiles, and for annihilating and decaying dark matter. In all cases, we show numerical results along with the analytic approximations (for cases 1 and 2 from Section \ref{sec:newform2}). We plot the ratios (rather than actual fluxes) since they are independent of the antiproton injection spectrum, dark matter annihilation cross section or decay width, and other such parameters, and highlight the variation from adopting the different propagation models. 

Since the diffusion coefficient increases faster with energy in the conventional formalism, antiproton flux decreases more rapidly, leading to a steady rise in the ratio of fluxes at high energies in all panels of  Figure~\ref{fig:result1}. The downward sloping behavior of the ratios at energies lower than $\sim$30 GeV is due to the fact that the convective wind term becomes important at these energies, where it has a comparatively stronger effect in the conventional formalism. In all cases, the calculated antiproton fluxes between 15 to 5000 GeV in the two models are quite close to each other, differing at most by about 25\%. 
The difference is smaller than the systematic uncertainty, which can be estimated as the difference between the results of applying the conventional formalism with the MIN, MED and MAX parameters (see, for example, fig.~7 in Ref.~\cite{winodm}). Thus, effects of position-dependent diffusion coefficient do not yet need to be included in deriving constraints on dark matter models from the antiproton flux measurements as long as the present uncertainties on the galactic propagation parameters are fully taken into account. On the other hand, the model with exponentially changing diffusion coefficient almost certainly better reflects the correct physics than the conventional one, 
and the solutions we obtained are almost as simple as the conventional ones, especially at high energies where the wind term can be ignored and analytic approximations can be used. We therefore advocate using it as a ``benchmark" propagation model in dark matter studies.

\subsection{Effect on PAMELA Constraints}

The PAMELA experiment recently measured the cosmic ray antiproton flux and antiproton to proton ratio up to $\sim 180$~GeV~\cite{pamela2}, extending its earlier results~\cite{pamela}. Since no excess of antiprotons over what is expected from conventional astrophysics is evident, these measurements can be used to place bounds on properties of dark matter; this has been done in several earlier works (see, for instance,~\cite{winodm, miimplications,  constraintsitalian, dmseeslight, newpamela}). 
As an illustration of our results, we recalculate these bounds using the propagation model of Section~\ref{sec:newform2}. We focus on a stable WIMP dark matter pair-annihilating into $W^+W^-$ pairs, such as a wino-like neutralino in the MSSM, or the lightest T-odd particle of the Littlest Higgs model with T-parity~\cite{LHDM}. Antiprotons are produced in hadronic decays of the $W$ bosons. 

We use the solution in Eqs.~\leqn{newsolexp}--\leqn{more_defs}, and set the convective wind term to zero. The source term $q(E, \textbf{x})$ can be written as
\beq
q(T,\textbf{x})=\frac{1}{2}\left(\frac{\rho(\textbf{x})}{m_{\chi}}\right)^2\langle\sigma v\rangle\left(\frac{dN_{\bar{p}}}{dT}\right)\,,
\eeq{source}
where $T=E-m_{\bar{p}}$ is the kinetic energy of the antiproton, and the fragmentation function $dN_{\bar{p}}/dT$ can be parameterized as~\cite{fragmentation}
\beq
\frac{dN_{\bar{p}}}{dx}= (p_1x^{p_3}+p_2|\log_{10}x|^{p_4})^{-1}\,,
\eeq{frag}
where $x=T/m_{\chi}$. This parameterization fits the results of the PYTHIA Monte-Carlo code \cite{pythia}. The parameters $p_i$ depend on the WIMP mass $m_{\chi}$ and, for annihilation into $W^+W^-$, can be written as \cite{fragmentation}
\beqa
p_1&=&(306.0m_{\chi}^{0.28}+7.2\times10^{-4}m_{\chi}^{2.25})^{-1},\CR
p_2&=&(2.32m_{\chi}^{0.05})^{-1},\CR
p_3&=&(-8.5m_{\chi}^{-0.31})^{-1},\CR
p_4&=&(-0.39m_{\chi}^{-0.17}-2.0\times10^{-2}m_{\chi}^{0.23})^{-1}.
\eeqa{parameters} 
This parameterization is valid for the WIMP mass in the range 50 GeV to 5 TeV. 

\begin{figure}[t]
\centering
\begin{tabular}{cc}
\includegraphics[width=3.2in,height=2.5in]{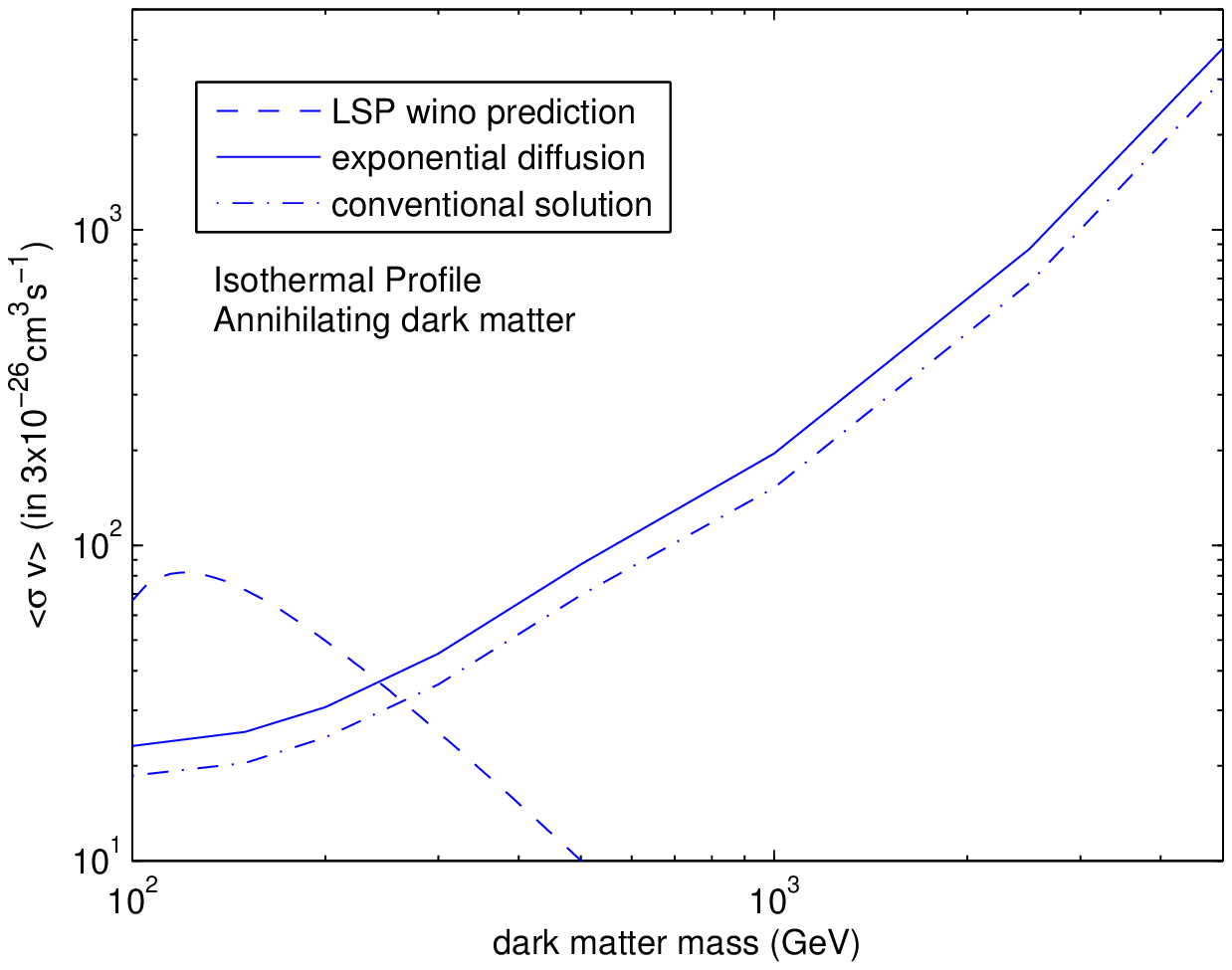}&
\includegraphics[width=3.2in,height=2.5in]{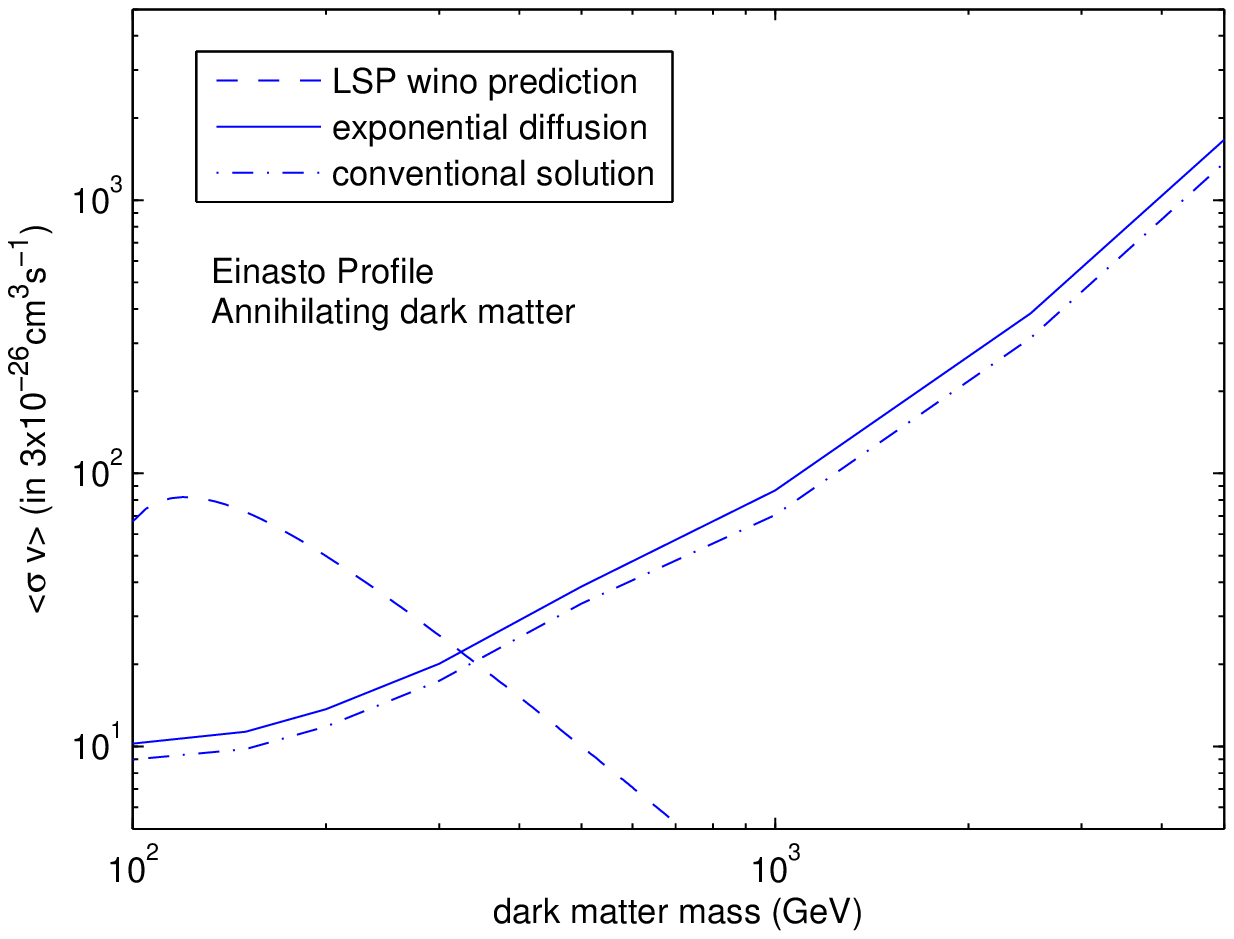}\\
\end{tabular}
\caption{Upper limits on annihilation cross section as a function of $m_{\chi}$ for $\chi\chi\rightarrow W^+W^-$ for isothermal (left) and Einasto (right) profiles. Solid (dot-dashed) curves denote limits from using the propagation model with exponentially increasing (constant) diffusion coefficient. The theoretical prediction for s-wave LSP wino annihilation is also given (dashed curve).}
\label{fig:constraints}
\end{figure}

To obtain bounds, we fit to the PAMELA antiproton flux data presented in~\cite{pamela2}, conservatively assuming zero antiproton background from astrophysical sources. Figure~\ref{fig:constraints} 
shows the bound on the annihilation cross section $\langle \sigma v \rangle$, obtained by requiring that the signal not exceed the measured flux in any of the energy bins by more than $2\sigma$ (statistical error only). The corresponding constraints calculated within the conventional propagation formalism (MED propagation model) are also shown; these are in agreement with similar limits calculated in~\cite{newpamela}. The model with the exponentially increasing diffusion coefficient is found to relax the constraints by $\sim 20\%$ for both isothermal and Einasto profiles. In most cases, the bound is enforced by the flux in the highest energy bin, at 100-180 GeV; Figure~\ref{fig:result1} indeed shows that there is a difference of about 20\% in the antiproton flux calculated in these two models. 

We also plot, for comparison, the theoretically predicted s-wave cross section for the MSSM wino annihilating into $W^+W^-$ \cite{theorycs}
\beq
\langle\sigma v\rangle \approx \frac{1}{m_{\chi}^2}\frac{0.65^4}{2\pi}\frac{(1-x)^{3/2}}{(2-x)^2}\,,
\eeq{csformula}
where $x=m_W^2/m_{\chi}^2$. The conventional propagation model rules out wino-like neutralinos lighter than $\sim 260$ GeV (for an isothermal profile); the model with the exponentially increasing diffusion coefficient slightly weakens this lower bound to $\sim 240$ GeV. As mentioned earlier, these results should be taken with a grain of salt, since in the conventional formalism calculation we fixed the parameters corresponding to the MED propagation model; including the uncertainties on these parameters would result in a band of predictions encompassing those of the model with exponential diffusion coefficient. 

\section{Discussion}
\label{sec:conc}

In this paper, we extended the conventional approach to calculation of the antiproton flux from dark matter annihilation/decay to allow for the possibility of a position-dependent diffusion coefficient. We studied two models, the three-zone model where the diffusion coefficient jumps on the boundary between the ``diffusion" and ``free-propagation" zones, and a model with the diffusion coefficient growing smoothly (exponentially) with distance away from the disk. In the first model, we found an analytic solution and showed that in the limit of infinite diffusion coefficient in the free-propagation zone the flux on Earth is not modified by the sources in that zone due to perfect reflection of antiprotons from the zone boundary. This seems to justify the conventional calculation even in situations when there are sources outside the diffusion zone, as is common in dark matter studies. On the other hand, the three-zone model is a rather crude approximation, since magnetic fields, and with them the diffusion coefficient, are expected to vary smoothly with distance from the disk. Such smooth variation was incorporated in our second model, which assumed an exponentially growing diffusion coefficient, which has been shown to produce consistent fits to conventional astrophysical cosmic ray observables in Ref.~\cite{exponentialk}. We found numerical solutions as well as analytic approximations valid at high antiproton energies for this model. We found that the resulting fluxes on Earth differ from the predictions of the conventional calculation with the MED propagation model by at most about 25\%.
Such deviations are well within the numerous uncertainties inherent in the calculation, hence the use of the conventional approach is justified, at least at present, in computing the antiproton fluxes from dark matter annihilation/decay and using them to place bounds on dark matter models. 
Nevertheless, since the model with exponentially growing diffusion coefficient almost certainly captures the correct physics of charged particle propagation in the galaxy better than the conventional one, and since the solutions we obtained in this model -- especially the analytic solutions, which are accurate at high energies -- are essentially as simple as the conventional ones, we advocate the use of this model as a benchmark for dark matter studies. 

\vskip0.5cm

\noindent{\large \bf Acknowledgments} 
\vskip0.3cm

We thank Joakim Edsj\"{o} and Jesse Thaler for useful discussions. B.S. would like to thank the Oskar Klein Center at Stockholm University, where part of this work was done, for their hospitality. This research is supported by the U.S. National Science Foundation through grant PHY-0757868 and CAREER grant No. PHY-0844667.

\end{document}